\documentclass[12pt]{article}

\usepackage{amsfonts}
\usepackage{epsfig}
\usepackage{graphicx}
\usepackage{latexsym}    
\usepackage{amsmath}
\usepackage{amssymb}
\usepackage{hyperref}

\newtheorem{lemma}{Lemma}

\newtheorem{theor}{\large\bf Theorem}

\newcommand{\qed}{{\hfill $\Box$}}

\def\FF{\hbox to 8.33887pt{\rm I\hskip-1.8pt F}}
\def\NN{\hbox to 9.3111pt{\rm I\hskip-1.8pt N}}
\def\PP{\hbox to 8.61664pt{\rm I\hskip-1.8pt P}}
\def\QQ{\rlap {\raise 0.4ex \hbox{$\scriptscriptstyle |$}}
{\hskip -4.5pt Q}}
\def\RR{\hbox to 9.1722pt{\rm I\hskip-1.8pt R}}
\def\ZZ{\hbox to 8.2222pt{\rm Z\hskip-4pt \rm Z}}

\newcommand{\mb}[1]{\mbox{$#1$}}
\newcommand{\be}{\begin{equation}}
\newcommand{\ee}{\end{equation}}
\newcommand{\bqa}{\begin{eqnarray}}
\newcommand{\eqa}{\end{eqnarray}}
\newcommand{\ba}{\begin{array}}
\newcommand{\ea}{\end{array}}

\newcommand{\no}{\nonumber}

\newcommand{\lp}{\left (}
\newcommand{\rp}{\right )}

\newcommand{\al}{\alpha}
\newcommand{\bt}{\beta}

\newcommand{\de}{\delta}

\newcommand{\vep}{\varepsilon}

\newcommand{\la}{\lambda}

\newcommand{\La}{\Lambda}

\newcommand{\De}{\Delta}

\newcommand{\bpsi}{\bar{\psi}}

\begin{document}

\title{Parametric Cutoffs for Interacting Fermi Liquids}

\author{M. Disertori$^1$, J. Magnen$^2$ and V. Rivasseau$^3$\\
1) Laboratoire de Math\'ematiques Rapha\"el Salem\\ CNRS UMR 6085,
Universit\'e de Rouen, 76801 Rouen Cedex\\
2) Centre de Physique Th\'eorique, CNRS UMR 7644,\\
Ecole Polytechnique F-91128 Palaiseau Cedex, France\\
3) Laboratoire de Physique Th\'eorique, CNRS UMR 8627,\\ 
Universit\'e Paris XI,  F-91405 Orsay Cedex, France}

\maketitle 

\begin{abstract}

This paper is a sequel to \cite{DMR}.
We introduce a new multiscale decomposition of the 
Fermi propagator based on its parametric representation. 
We prove that the corresponding sliced propagator obeys
the same direct space bounds than the decomposition used
in \cite{DMR}. Therefore the non perturbative bounds on completely convergent
contributions of \cite{DMR} still hold. In addition the new slicing better  
preserves momenta, hence should become an important new technical tool 
for the rigorous analysis of condensed matter systems.
In particular it should allow to  complete the
proof that a three dimensional interacting system of Fermions 
with spherical Fermi surface is a Fermi liquid in the 
sense of Salmhofer's criterion. 
 
\end{abstract}

\section{Introduction}

Interacting Fermi liquid theory is not valid down to zero temperature. 
Below some critical temperature the quasi-particles with momenta near the Fermi surface
bound into Cooper pairs. This generic phenomenon goes under 
the name of Kohn-Luttinger instabilities.
Hence the mathematical definition of 
Fermi liquid behavior is not obvious.

There are essentially two main ways to block the formation of Cooper
pairs, namely to increase temperature or magnetic field.

With a generic strong magnetic field, parity invariance of the Fermi 
surface is broken
and a true discontinuity at a well-defined Fermi surface may be 
proved mathematically.
This is the road followed by Feldman, Kn\"orrer and Trubowitz in the
impressive series of papers \cite{FKT}, in which they
proved two dimensional Fermi liquid behavior at zero temperature
for sufficiently convex and regular parity-breaking Fermi surfaces. 

Magnetic fields responsible for parity breaking 
are also the source of the quantum Hall effect.
The rigorous treatment of this effect could 
require a non-commutative formulation of renormalizaton group 
\cite{PPR} and a suitable generalization of the 
parametric cutoffs of the present paper.

A criterion to characterize Fermi liquid behavior without breaking
parity  has been proposed in \cite{Sal}.
Salmhofer remarked that staying in a domain $ |\lambda \log T| \le K$, 
where $\lambda$ 
is the coupling constant and $T$ is the temperature, avoids Cooper pairing
and Kohn-Luttinger singularities. 
Therefore after mass renormalization of the two point function 
Schwinger functions should be analytic 
in $\lambda$ in such a domain $ \vert \lambda \log T| \le K$. Salmhofer criterion states that for Fermi liquids
the self-energy and its first and second momentum derivatives remain 
uniformly bounded in such a domain. 

Fermionic models in one dimension are Luttinger liquids 
 \cite{BG,BGPS,BM} and they do not obey the Salmhofer criterion.
In two dimensions it has been proved that interacting
Fermion systems with a circular \cite{DR1} or 
approximately circular  \cite{BGM} Fermi surface obey this criterion.
 In contrast the Hubbard model at half filling,
which has a square Fermi surface, violates the criterion and its
self-energy behaves as a Luttinger liquid with logarithmic corrections  \cite{Hub}.
Hence Salmhofer criterion effectively distinguishes Fermi-like liquids
from Luttinger-like liquids in two dimensions.
All these results rely on the special momentum conservation 
rules of interacting Fermi systems in two dimensions. 

In three dimensions Fermi liquid behavior is generically expected
but momentum conservation rules allow for
non planar vertices and the two-dimensional methods
do not extend to the constructive level. The only existing constructive method 
has been pioneered in  \cite{MR} and further developped in  \cite{DMR}.
 It relies on a direct space decomposition of the propagator combined with 
cluster expansions and Hadamard's inequalities.
This is a bit surprising for a constructive Fermionic problem,
which can usually be treated with Gram's inequalities and no cluster expansions
\cite{Les,AR}. For a recent pedagogical introduction to these questions
and further explanations see \cite{Riva}.

It was proved in \cite{DMR} that the sum of all convergent contributions
is indeed analytic in a Salmhofer domain.
However the mass renormalization (which in this context can be interpreted 
as a change of the Fermi surface radius) was not performed, 
and Salmhofer's criterion was not checked,  
because the cutoffs used on the propagator 
did not conserve momentum, hence  the computation of the self-energy (i.e. the one-particle  irreducible (1PI)  amputated two point function) was not automatic
in this formalism. Extracting the 1PI two point function 
would have required a sequence  of Mayer expansions to remove hard-core 
constraints in the cluster expansion.

The situation remained in this incomplete stage for a decade. 
In this paper we define a new slice decomposition of the propagator at finite temperature which approximately conserves momentum.
Our main result is Theorem 1 in Section 4 which states that this new slicing obeys the spatial bounds of the former slicing used in  \cite{DMR}. Its proof relies on a saddle point analysis with rigorous control of the remainder terms.

As a consequence the bounds of \cite{DMR} 
on convergent contributions which do not require
mass renormalization also hold for this new decomposition, but
mass renormalization of the two point divergent subgraphs should become much easier. Indeed momentum preserving cutoffs have the nice property that two point subgraphs made of higher slices than their external legs are automatically one particle irreducible. This was the key to simplify their renormalization and 
to prove Salmhofer's bounds in all previous works \cite{DR1,BGM,Hub}. 

It is therefore likely that using this new propagator slicing the program of \cite{DMR} can be completed, although the mass renormalization and 
the complete proof of Salmhofer's criterion remain beyond 
the scope of the present paper.

\section{The model}

The model is the isotropic jellium in three spatial dimensions
with a local four point interaction considered in \cite{DMR}.
We recall for completeness the corresponding notations
and conventions.

\subsection{Free propagator}

Using the Matsubara formalism,
the propagator in Fourier space $\hat{C}$  is equal to:

\begin{equation}
\hat{C}_{ab} (k_0, k) = \de_{ab} \frac{1}{ik_0-e(k)},
\quad \quad e(k)= \frac{k^2}{2m}-\mu \ ,
\label{prop}
\end{equation}
where $a,b \in \{\uparrow,\downarrow\}$ are the
spin indices. 
The vector $ k$ in (\ref{prop}) is three-dimensional. 
The parameters $m$ and $\mu$ correspond to the effective mass and to
the chemical potential (which fixes the Fermi energy).
To simplify notation we put $2m= \mu=1$, so that
$e(k)=  k^2-1$. The corresponding direct space propagator at temperature
 $T$ and position $(t, x)$
(where $x$ is the three dimensional spatial component) is 
\begin{equation}
C_{ab}(t,x) =\frac{T}{(2\pi)^3}\; \sum_{k_0 } \; \int d^3k\; e^{-ik_0 t +ikx}\;
\hat{C}_{ab}(k_0, k) \ ,
\label{tfprop}\end{equation}
and is antiperiodic in the variable
$t$ with antiperiod $\frac{1}{T}$. This means that
\begin{equation}
\hat{C}(k_0,k)= \frac{1}{2}\int_{-\frac{1}{T}}^{\frac{1}{T}} dt
\int d^3x \;e^{+ik_0 t -ikx}\; C(t,x)
\end{equation}
is not zero only  for   discrete
values (called the Matsubara frequencies) :
\begin{equation}
k_0 =   (2n+1) \pi T \ , \quad n \in \ZZ \ , 
\label{discretized}
\end{equation}
where we take $ /\!\!\!{\rm h} =k =1$. Remark that only
odd frequencies appear, because of  antiperiodicity, 
hence $\vert k_0\vert  \geq \pi T$
so that the temperature acts like an effective infrared cutoff.

The notation $\sum_{k_0}$ in (\ref{tfprop})
means really the discrete sum over the integer
$n$ in (\ref{discretized})\footnote{When $T \to 0$,  $k_0$
becomes a continuous variable, the discrete sum becomes an
integral  $T\sum_{k_{0}}\rightarrow \frac{1}{2\pi }\int dk_{0} $, 
and the corresponding propagator
 $C_{0}(k_0,k)$ becomes singular
on the Fermi surface
defined by $k_0=0$ and $|k|=1$.}.
To simplify notations we write:
\begin{equation}
\int d^4k \; \equiv \; T\sum_{k_0} \int d^3k
\quad , \quad 
\int d^4x \; \equiv \; \frac 12
\int_{-1/T}^{1/T}dt \int d^3x \ . \label{convention}
\end{equation}

\subsection{Propagator with ultraviolet cutoff}

We remember that  we can add a continuous ultraviolet cut-off
(at a fixed scale $\La_{u}=1$) to the propagator (\ref{prop}).  For convenience
we introduced this cutoff both on spatial and on Matsubara frequencies; 
indeed the 
Matsubara cutoff could be lifted with little additional work.
The propagator (\ref{prop}) equipped with this cut-off is called
$C^{u}$ and   is defined as:
\begin{equation}\label{Cu}
\hat{C}^{u}(k_0, k) :=   \frac{e^{-[k_0^2 + (k^2 -1)^2]}}{ik_0 + (k^2-1)} \ . 
\end{equation}
Note that in previous works we used a compact support function for 
this ultraviolet cutoff.  
Here we use an exponential because it is better adapted 
to the parametric representation that we shall use.

\subsection{Partition function}

Finally we introduce  the local four point interaction 
\begin{equation}
I(\psi,\bpsi) = \la \int_\La d^4x \; (\bar{\psi}_\uparrow \psi_\uparrow)
(\bar{\psi}_\downarrow \psi_\downarrow) =
 \la \int_\La d^4x \; \prod_{c=1}^4 \psi_c \ ,
\end{equation}
where $\psi_c$ is defined as: 
\begin{equation}
\psi_1= \bar{\psi}_{\uparrow} \quad \psi_2= \psi_{\uparrow}\quad 
 \psi_3= \bar{\psi}_{\downarrow} \quad  \psi_4= \psi_{\downarrow} 
\label{psic}\end{equation}
The partition function is then defined as
\bqa
Z_\La^{u} &=& \int
d\mu_{C^{u}}(\psi,\bpsi)
e^{I(\psi,\bpsi)}
= \sum_{n=0}^\infty \frac{1}{n!} \int
d\mu_{C^{u}}(\psi,\bpsi)
I(\psi,\bpsi)^n\no\\
&=& \sum_{n=0}^\infty \frac{1}{n!} \int
d\mu_{C^{u}}(\psi,\bpsi)
\prod_{v\in V} I_v(\psi,\bpsi)
\eqa
where $V$ is the set of $n$ vertices and $I_v(\psi,\bpsi)$ denotes
the local interaction at  vertex $v$. 

In order to perform a multiscale analysis we need to
introduce a slice decomposition over fields. This 
corresponds to  introduce a  slice decomposition on the free
propagator (with UV cutoff) $C^{u}$. 

\section{The propagator}

In \cite{DMR} we introduced a multiscale analysis
directly on position space. It is more convenient to introduce a 
new scale decomposition 
which is compatible with momentum conservation. This conservation is 
indeed useful to control renormalization of two point functions.

This new decomposition is the main technical innovation of this paper 
with respect to \cite{DMR}. It cuts slices on the integration range of 
the Schwinger parameter: 
this a good compromise between $x$ and $p$ space slicing. 

\subsection{Schwinger representation  of the propagator}

\begin{lemma}\label{l:param}
The propagator (with UV cutoff) $C^{u}$ defined above \eqref{Cu} can be
written as
\begin{align}
\hat{C}(k_0, \vec{k})&= 
\int_1^{\infty} d\alpha \int_{\mathbb{R}} d\beta  \ 
\hat{F}(\alpha,\beta, k_0, \vec{k})
 \label{prep}  \\
C(t,\vec{x})  &=  \int_1^{\infty} d\alpha \int_{\mathbb{R}} d\beta\ 
F(\alpha,\beta,t, \vec{x})  \label{xrep}
\end{align}
where
\begin{equation}\label{prep0}
\hat{F}(\alpha,\beta, k_0, \vec{k}) =  [ -ik_0 + (k^2-1)] 
 \frac{1 }{2\sqrt{\pi \al}} e^{-\alpha k_0^2 }
 \ e^{- \frac{\beta^2}{4 \alpha} + i \beta (k^2 -1)}
\end{equation}
\begin{align}
F (\alpha,\beta,t, \vec{x}) &=  - 
\tfrac{\sqrt{\pi}}{2(2\pi)^4} 
\ \tfrac{{\tilde I} (\alpha ,t) +i\bt I (\alpha,t) }{\alpha^2}  \  
e^{- \frac{\beta^2}{4 \alpha}} 
\int \ d^3k \ e^{+i k . x}\ e^{i \beta ( k^2 -1)} \label{xrep0}\\
&= - \tfrac{e^{i\frac{3\pi}{4}}}{2(4\pi)^2} \ 
\tfrac{{\tilde I} (\alpha ,t) +i\bt I (\alpha,t ) }{\alpha^2 \beta^{3/2}}  \ 
 e^{- \frac{\beta^2}{4 \alpha}}  e^{-iB (\beta,x)}\ , \label{xrep1}
\end{align}  
we do not write the  $u$ index for simplicity and we defined
\begin{align}
 B (\beta ,x) =  \lp \bt + \frac{x^2}{4\bt}\rp
\label{Bdef}
\\
I(\al,t)  =   (\sqrt{\al}T) \sum_{k_0}  e^{-ik_0 t } e^{-\al k_0^2} & 
\qquad 
{\tilde I} (\alpha ,t) = - 2\al\ \partial_t I(\al,t)\ .\label{Idef}
\end{align}
Moreover for  $\al< T^{-2}$ and for any $p>0$, 
\begin{equation}
|I| \ \leq   \   \frac{K_p}{\left (1+ \frac{f(t)}{\sqrt{\al}} \right )^p  } 
\ , \qquad
{\tilde I} \ \leq \ \sqrt{\al}  \frac{K_p}{\left (1+ \frac{f(t)}{\sqrt{\al}}
 \right )^p  }
\label{I1b}
\end{equation}
and when  $\al \geq T^{-2}$
\begin{equation}
|I| \  \leq   \frac{c}{1+T f(t)}\leq K   \ , \qquad
{\tilde I} \leq \frac{c}{T}  \frac{1}{1+T f(t)} \leq \frac{K}{T}\ ,
\label{I2b}
\end{equation}
where $K_p$ is a constant depending only of $p$, $K$ and $c$ are
constants  and 
$f(t)$ is defined by 
\begin{equation}
f(t) = \bigg| \frac{\sin (2\pi T t )}{2\pi T} \bigg|\ , \qquad t\in 
\left[ -\frac{1}{T},\frac{1}{T}\right]
\label{f}
\end{equation}
\end{lemma}

\paragraph{Remark 1} 
Note that $f(t)\leq 1/T$ hence the bound in \eqref{I2b}.
\paragraph{Remark 2}  The function $I (\alpha,t)$ 
is a discrete Fourier transform. 
In the continuum limit we have
\begin{equation}
I (\alpha ,t)  = \frac{1}{2 \sqrt{\pi}} e^{- \frac{t^2}{4\al}} \qquad 
{\tilde I} (\alpha ,t) = \frac{1}{2 \sqrt{\pi}} t\ e^{- \frac{t^2}{4\al}}
\label{c-lim}\end{equation}
so $t$ cannot be larger than $\sqrt{\al}$. This should be true 
also for $T$ finite, but instead of exponential  
we may expect only polynomial decay \eqref{I1b}. Moreover, as $|t|\leq 1/T$ we 
get a decay not directly for $\vert t \vert $ but for $f(t)$ \eqref{f},
which is what we otained also for the full
propagator in $x$ space (see \cite{DMR}, section II.3).
As in that case, the proof will be based on integration
by parts.

\paragraph{Proof of the first part: \eqref{prep} and  \eqref{xrep} }
The Schwinger representation of the propagator (with its UV cutoff) is
\begin{align}
\hat{C}(k_0, \vec{k}) &= [ -ik_0 + (k^2-1)] 
\int_1^{\infty} d\alpha  \ e^{-\alpha[k_0^2 + ( k^2 -1)^2 ]}
\end{align}
We rewrite the quartic term in the exponent as a Gaussian integral
\begin{equation}
 e^{-\alpha ( k^2 -1)^2 }  = \frac{1}{2\sqrt{\pi \al}}
 \int_{\mathbb{R}} d\beta \ e^{- \frac{\beta^2}{4 \alpha} + i 
\beta ( k^2 -1)}\ ,
\end{equation}
 so the propagator becomes
\begin{equation}
\hat{C}(k_0, k) =  [ -ik_0 + (k^2-1)] 
\int_1^{\infty} d\alpha   \ 
 \frac{ e^{-\alpha k_0^2 }}{2\sqrt{\pi \al}}
 \int d\beta \ e^{- \frac{\beta^2}{4 \alpha} + i \beta (k^2 -1)}.
\end{equation}
In order to get the $x$ behavior we need to take the Fourier transform 
\be
C(t,x) = \frac{1}{(2\pi)^3}  \int \ d^4k \ e^{-ik_0t + ik.x} \hat{C}(k_0, k)
\end{equation}
Now using the following relations\footnote{To prove the first relation: 
the linear term can be eliminated by
a real translation; the remaining integral can be computed on the 
positive real semi-axis by
rotating the contour in the complex plane by an angle $\pi/4$.}
\begin{align}
&\int \ d^3k \ e^{+i k . x}\ e^{i \beta ( k^2 -1)} \ = \  
e^{-i(\bt +\frac{ x^2}{4\bt})} 
 \left ( e^{i\frac{\pi}{4}} \sqrt{\frac{\pi}{\bt}} \right )^{3}
 \cr
& [ -ik_0 + (k^2-1)]  e^{-ik_0 t + i \beta ( k^2 -1)} \ =\  
[ \partial_t  - i \partial_\bt ]  
 e^{-ik_0 t + i \beta ( k^2 -1)}
\end{align}
and the definition \eqref{Idef} we get 
\begin{align}
C(t,x) =&  \frac{1}{(2\pi)^3}
\int_1^{\infty}  \tfrac{ d\alpha}{2\al \sqrt{\pi }} \int_{\mathbb{R}} d\beta\  
 \ e^{- \frac{\beta^2}{4 \alpha}} 
  [ \partial_t  - i \partial_\bt ]  
\cr & \hskip 3cm 
\left [  I(\al,t)
 e^{-i(\bt +\frac{x ^2}{4\bt})} 
 \left ( e^{i\frac{\pi}{4}} \sqrt{\frac{\pi}{\bt}} \right )^{3}\right ]\cr
 =&  \frac{1}{(4\pi)^2} 
 \int_1^{\infty}  \frac{d\al }{\al}\  \int_{\mathbb{R}} 
 \frac{d\beta\ }{\bt^{3/2}}  \ 
   e^{-i (-3\frac{\pi}{4} +\bt + \frac{x^2}{4\bt} )} 
\left [ \partial_t I +i I\partial_\bt \right ] e^{- \frac{\beta^2}{4 \alpha}} 
  \cr
=&   - \frac{e^{i\frac{3\pi}{4}}}{ (4\pi)^2} 
 \int_1^{\infty}  \frac{d\al }{\al}\  \int_{\mathbb{R}} 
 \frac{d\beta\ }{\bt^{3/2}}  \ 
   e^{-i (\bt + \frac{x^2}{4\bt} )} 
 e^{- \frac{\beta^2}{4 \alpha}} \  \lp\tfrac{{\tilde I}+i\bt I}{2\al}\rp \cr
 &= -\frac{e^{i\frac{3\pi}{4}}}{2(4\pi)^2} 
\int_1^{\infty} d\alpha \int_{\mathbb{R}} d\beta\ 
\frac{{\tilde I}+i\bt I }{\alpha^2 \beta^{3/2}}  
e^{-\frac{\beta^2}{4 \alpha}} \   e^{-iB} \label{c}
\end{align}
where in the second line we applied integration by parts with respect 
to $\bt$, and the
definitions \eqref{Idef} for $\tilde{I}$ and \eqref{Bdef} for $A,B$.
Note that 
\begin{align}
|I| &\leq \ \frac{\sqrt{\al}}{(2\pi)}  \int\ dk_0\    e^{-\al k_0^2} = 
\frac{1}{2 \sqrt{\pi}}\cr
|{\tilde I}|  &\leq \ 2 \frac{\al^{3/2}}{(2\pi)}   
\int\ dk_0  \ |k_0| e^{-\al k_0^2}
= \frac{\sqrt{\al}}{\pi}\ .
\end{align}
Therefore the final expression is integrable separately in $\al$ and $\bt$ and 
the integration order no longer matters.
This completes the proof of \eqref{prep} and \eqref{xrep}. If we do not 
perform explicitely the Fourier transform with respect to $\vec{k}$ we get
\eqref{xrep0}.
\qed 

\paragraph{Proof of the second part: \eqref{I1b} and \eqref{I2b} }

It remains to prove the decay for $I$ and $\tilde{I}$. This is done using 
integration by parts.

From now on $K_p$ is a generic name for a constant that depends only on $p$,
and we may use simplification such as $K_p K_p = K_p$, $const . K_p = K_p$
(but of course only finitely many times...).

The key identity is 
\begin{equation}
\left [1+\frac{f(t)}{\sqrt{\al}}\right] 
e^{-ik_0 t} = \left [1+ i\vep(t) \frac{1}{\sqrt{\al}} \frac{\De}{\De k_0} 
\right ]  e^{-ik_0 t} \label{tident}\end{equation}
where $\vep(t)$ is the sign of $\sin \lp 2\pi Tt\rp$ and the 
discretized derivative  $\frac{\De}{\De k_0}$
on a function $F(k_0)$ is defined by
\begin{equation}
\frac{\De}{\De k_0}  F(k_0) = \frac{1}{4\pi T} \left [
F(k_0+2\pi T) - F(k_0-2\pi T)\right ] .
\end{equation}
Let us consider first the case $p=1$, then we 
apply this identity inside $I$ only once. Performing integration 
by parts we get 
\begin{eqnarray}
I (\alpha ,t) 
&=& \sqrt{\al} T \frac{1}{\left (1 + \tfrac{f(t)}{\sqrt{\al}}\right ) } 
\sum_{k_0} e^{-i t k_0} 
\left [1- i\vep(t) \frac{1}{\sqrt{\al}} \frac{\De}{\De k_0} \right ] 
e^{- \al k_0^2}  \\
&=&  \sqrt{\al} T \tfrac{1}{\left (1 + \tfrac{f(t)}{\sqrt{\al}}\right )} 
\sum_{k_0} e^{-i t k_0} 
\left [1+ i\vep(t)  \tfrac{\sinh 4\pi T \al k_0}{2\pi T\sqrt{\al}} 
e^{-\al (2\pi T)^2}\right ] 
e^{- \al k_0^2} .\nonumber
\end{eqnarray}
Now inserting absolute values we have
\begin{eqnarray}
 |I| \left (1 + \tfrac{f(t)}{\sqrt{\al}}\right ) \ & \leq & \   
 \sqrt{\al} T  \sum_{k_0} 
\left [1+  \tfrac{|\sinh 4\pi T \al k_0|}{2\pi T\sqrt{\al}} 
e^{-\al (2\pi T)^2}\right ] 
e^{- \al k_0^2}\\
\nonumber
 &\leq  & 2 \sqrt{\al} T  \sum_{k_0>0} 
\left [1+  \tfrac{\sinh (4\pi T \al k_0)}{2\pi T\sqrt{\al}} 
e^{-\al (2\pi T)^2}\right ] e^{- \al k_0^2}\\
 &\leq &  2\sqrt{\al} T  \sum_{k_0>0} 
\left [1+  2\sqrt{\al} k_0\  e^{4\pi T \al k_0}
e^{-\al (2\pi T)^2}\right ] e^{- \al k_0^2} 
 \ \leq \ const, \nonumber
\end{eqnarray}
where in the last line we used that $\sinh x \leq x\ e^x $ for all  
$x\geq 0\ $.
For $p>1$ we must apply \eqref{tident} $p$ times:
\begin{align}
& \quad I (\alpha ,t)\  \left (1 + \tfrac{f(t)}{\sqrt{\al}}\right )^p = 
\sqrt{\al} 
T \sum_{k_0} e^{-i t k_0} 
\left [1- i\vep(t) \frac{1}{\sqrt{\al}} \frac{\De}{\De k_0} \right ]^p 
e^{- \al k_0^2}.
\end{align}
Each new derivative
\begin{itemize}
\item either extracts a new $\left [1+ i\vep(t)  
\tfrac{\sinh 4\pi T \al k_0}{2\pi T\sqrt{\al}} e^{-\al (2\pi T)^2}\right ]$ 
factor from the exponential,
\item or applies to a factor derived before.
\end{itemize}
So we obtain a sum of terms of the following type
\begin{align}
& 1) \ \left (\tfrac{\sinh 4\pi T \al k_0}{2\pi T\sqrt{\al}}\right )^n 
e^{-n\al (2\pi T)^2} 
\ \leq 
(\sqrt{\al} k_0)^n e^{ n 4\pi T \al k_0} e^{-n\al (2\pi T)^2}
\quad {\rm for }\  n\leq p,\cr
& 2)\ \left (\tfrac{1}{\sqrt{\al}} \tfrac{\De}{\De k_0} \right )^{2n}  
\tfrac{\sinh 4\pi T \al k_0}{2\pi T\sqrt{\al}} \ = \ 
\sinh 4\pi T \al k_0\ 
\left (\tfrac{\sinh 8\pi^2 T^2 \al}{2\pi T\sqrt{\al}}\right )^{2n-2} 
 \left(\tfrac{\sinh 8\pi^2 T^2 \al}{2\pi T^2\al}\right ) \cr
 & \quad \quad \leq e^{4\pi T \al k_0}  const\quad  {\rm for }\ 2n\leq p ,\cr
 &3)\ \left (\frac{1}{\sqrt{\al}} \frac{\De}{\De k_0} \right )^{2n+1}  
\tfrac{\sinh 4\pi T \al k_0}{2\pi T\sqrt{\al}} \ = \ 
\cosh 4\pi T \al k_0\ 
\left(\tfrac{\sinh 8\pi^2 T^2 \al}{2\pi T\sqrt{\al}}\right )^{2n-1} 
 \left(\tfrac{\sinh 8\pi^2 T^2 \al}{2\pi T^2\al}\right ) \cr
 &\quad  \quad \leq e^{4\pi T \al k_0}  const\quad  {\rm for }\  2n+1 \leq p\ .
 \end{align}
The first term after summing over $k_0$ is bounded by 
$K_p \ e^{n(n-1) \al (2\pi T)^2}$. This
factor is bounded as $\al T^2\leq 1$. In the second and last terms the factors 
$\sinh(8\pi^2 T^2 \al)/(2\pi T^2\al)$ and 
$\sinh(8\pi^2 T^2 \al)/(2\pi T\sqrt{\al})$ are
bounded as long as $\al T^2\leq1$.
The same arguments hold for ${\tilde I}$. 

In the case  $\al T^2>1$ we cannot go beyond $p=1$ as this time the factors 
$e^{n(n-1) \al (2\pi T)^2} $
are not bounded. 
\qed

\subsection{Slice decomposition on the propagator}

After  introducing the Schwinger representation of the propagator 
we obtained \eqref{prep}-\eqref{xrep}.   
Now, fixing a positive number $M>1$ we want to cut out as usual
$j_m$ main RG slices following a geometric progression of ratio $M$, where
$j_m$ is defined as the temperature scale such that
$M^{j_m}\simeq  1/T$, more precisely
\begin{equation}
j_m =1+ Int\left [ \frac{\ln   \lp T^{-1} \rp }{\ln M}\right  ] 
\label{jM}
\end{equation}
where $Int$ means the integer part. 

\subsubsection{Heuristic analysis} Remark that $x$ enters only in the 
oscillating factor $e^{B}$ of the  $\bt$ integral.
For $|x|$ large this integral should be approximated by the region near 
the saddle points, where $\partial_\beta B=0$,
namely $|\bt | = |x|/2$. On the other hand, $t$ enters only in the $I$ and 
${\tilde I}$ factors. The $t$ dependence is controlled by the 
decay of these factors, which is in $\left (1+ f(t)/\sqrt{\al}\right )^{-p}$ 
for $\al\leq 1/T^2$, 
so the $\al$ integral  for $t$ large should be concentrated around 
$\al \gtrsim t^2$, and
in fact around $\alpha \simeq t^2$ (taking into account the $\al^{-2}$ which 
ensures convergence of the $\al$ integral). 

Recall that in  \cite{DMR} the decomposition was done in $x$ space, 
with as key relation defining the main slice:
\begin{equation}
M^{j-1} \leq (1+ |x|)^{\frac{3}{4}}  (1+|t|+ |x|)^{\frac{1}{4}} \leq M^j\ .
\end{equation}
For $|t|\leq |x|$ this relation gives $|x|\propto M^j, |t|\leq M^j$. For
$|t|> |x|$ an auxiliary decoupling $|t|\propto M^{j+k}$ was introduced.
Then $|x|\propto M^{j-k/3}$. We can mimick this slicing by observing that 
$\al\simeq t^2$, $|\bt |\simeq |x|$. Hence the slicing relations in 
parametric space should be
\begin{equation}
M^{2j-2} \leq (1+ \bt^2)^{\frac{3}{4}}  (\al+ \bt^2)^{\frac{1}{4}} \leq 
M^{2j}\ , \qquad \al \simeq M^{2j+2k}\ .
\end{equation}
However since stationary phase analysis should not be done with sharp boundary
to avoid large boundary terms, we have to introduce 
these relations through smooth rather than sharp cutoff functions 
in the parametric space.

\subsubsection{Notation} For any two real numbers $X$, $Y$ we will write 
$X\leq_{c} Y$ if there is a constant $1/10<C<10$ such that 
$X\leq C Y$. The same holds for  $X\geq_{c} Y$ and $X\simeq Y$.

\subsubsection{The slicing}

Motivated by this heuristic discussion we 
write the propagator $C=\sum_{j=0}^{j_{m}} C^{j}$,  where 
\begin{equation}
C^{j}(t,x) =
 \int_1^{\infty} d\alpha   \int_{\mathbb{R}}  d\beta \  
\chi^{j}\big(X_{\al\bt}\big)\  F (\alpha,\beta,t,\vec{x})
\end{equation}
where 
\begin{equation}
X_{\al\bt} =  (1+ \bt^2)^{\frac{3}{4}}  (\al+ \bt^2)^{\frac{1}{4}}, 
\end{equation}
and
\begin{align}
\chi_j(X) &= u\left ( \frac{X}{M^{2j}} \right ) -  u\left ( \frac{X}{M^{2j-2}}
\right )
\ \qquad
j_m > j> 0\cr
\chi_0(X) &= u\left ( X \right )\ ,\ \qquad j=0\cr
\chi_{j_m}(X) &= 1 -  u\left ( \frac{X}{M^{2j_m-2}} \right )\ , \ \qquad 
j=j_m
\end{align}
and $u(x)$ is a smooth function with compact support such that 
$u(x)=1$ for $0\leq x\leq 1$ and $u(x)=0$ for $x>2$.  
Note that this definition implies the following constraints:
\begin{align}
\chi_j(X_{\alpha \beta}) &\neq 0\quad  \Rightarrow\ 
\ \quad X_{\alpha \beta}\simeq M^{2j} \cr
\chi_0(X_{\alpha \beta} ) &   \neq 0\quad  \Rightarrow\ 
\ \quad X_{\alpha \beta} \leq_{c} 1 \cr
\chi_{j_m}(X_{\alpha \beta} ) &  \neq 0\quad  \Rightarrow\ 
\ \quad X_{\alpha \beta} \geq_{c} M^{2j_m}
\end{align}
where in the first line we took $j_m > j> 0$.  

\paragraph{Remark} Actually,  in \cite{DMR} we introduced $j_{m}+1$ scales,
but all the estimates we obtained remain valid with $j_{m}$ scales.
\medskip

Now, as in \cite{DMR} we distinguish two situations.
\begin{itemize}
\item If $\al \leq \bt^2$, then $\chi_j(X)\neq 0$ only for $\bt\simeq M^j$. 
For the last scale $j=j_m$ we will have to distinguish the case
$\al \leq 1/T^2 = M^{2j_m}$ and the case  $M^{2j_m}< \al \leq  \bt^2$.
\item If $\al \geq \bt^2$,  we have to add an auxiliary decomposition over
the possible size of $\al$.
\end{itemize}

\subsubsection{Auxiliary scales}

As in \cite{DMR} for each $j\leq j_m$ we add the decomposition
\begin{equation}
1 = \sum_{k=0}^{k_m(j)} \tilde\chi_{k}(\al)\ , 
\end{equation}
where for $k_m (j) >0$ we define
\begin{align}
\tilde\chi_k(\al) &= u\left ( \frac{\al}{M^{2j+2k}} \right ) 
-  u\left ( \frac{\al}{M^{2j+2k-2}} \right )\  \  {\rm for}\
k_m(j) > k\geq 1 \ ,\cr
\tilde \chi_0(\al) &= u\left ( \frac{\al}{M^{2j}} \right )\ , \qquad k=0\cr
\tilde \chi_{k_m(j)}(\al) &= 1 -  u\left ( \frac{\al}{M^{2j+2k_m(j)}} \right
)\ ,
\qquad k=k_m(j) \ ,
\end{align}
and for $k_m=0$ we have no decomposition:
\begin{equation}
\tilde\chi_0(\al) = 1\ .
\end{equation}
Finally, as in \cite{DMR} $k_{m}(j)$ is defined as
\begin{equation}
k_m(j) = {\rm min} \left [(j_m-j), 3j\right ]  .
\end{equation}
These definitions imply the following constraints on $\alpha$ (when 
$k_m (j)>0$):
\begin{align}
\tilde\chi_k(\al) &\neq 0 \quad  \Rightarrow\ 
\ \quad \al \simeq M^{2j+2k}  \ ,\cr
\tilde \chi_0(\al) &\neq 0 \quad  \Rightarrow\ 
\ \quad 1\leq \al\leq_{c}  M^{2j}\ , \cr
\tilde \chi_{k_m(j)}(\al) &\neq 0 \quad  \Rightarrow\ 
\ \quad \al\geq_{c} M^{2j+2k_m}
\ , 
\end{align}
where in the first line we take $k_m (j)> k> 0$.  

\paragraph{Remark} In \cite{DMR} the bound $k\leq 3j$ was
obtained observing that $f (t)^{1/4}\leq M^{j}$ for $j\leq j_{m}$,
while the bound $k\leq j_{m}-j$ was due to $f (t)\leq M^{j_{m}}$
(by definition of $f (t)$). Here the first bound is still valid
since $\alpha^{1/4} \leq M^{2j}$, but the second no longer holds since
$\alpha $ can take any value in $[1,\infty)$. Nevertheless we take
the same definition of $k_{m} (j)$ as in \cite{DMR} so that the
results we obtained there can be directly applied here.
\medskip

The slicing of the propagator is then
\begin{equation}
C=\sum_{j=0}^{j_m} \sum_{k=0}^{k_m(j)} C^{jk} \ ,
\end{equation}
where
\begin{equation}
 \label{subsli}
C^{jk}(t,x) = 
 \int_1^{\infty} d\alpha   \int_{\mathbb{R}}   d\beta \  \chi_{jk}(\al,\bt)
\  F (\alpha,\beta,t,\vec{x})\ ,
\end{equation}
$F (\alpha,\beta,t,\vec{x})$ was introduced in 
\eqref{xrep0}-\eqref{xrep1} and we defined
\begin{equation}
\chi_{jk}(\al,\bt)=\chi^{j}(X_{\alpha \beta})\ 
\tilde\chi^k(\al) \ .
\label{eq:chijkdef}
\end{equation}
Note that this slicing selects $\al \geq \bt^2$ 
when $k>0$ and $\al\leq \bt^2$ when $k=0$. This is proved in the
following lemma.

\begin{lemma}
If $0<j<j_m$ then $k_{m}>0$ and: 
\begin{itemize}
\item for all $k_{m}\geq k>0$ we have 
$\chi_{jk}\neq 0$ $\Rightarrow$ $\al\geq_{c} \bt^2$ and
\item for $k=0$  we have $\chi_{jk}\neq 0$ $\Rightarrow$ $\al\leq_{c} \bt^2$.
\end{itemize}
\end{lemma}

\paragraph{Proof}
Since  $0<j<j_m$, $k_{m}>0$ by definition and  $\chi_{j}\neq 0$ implies
$X_{\alpha \beta }\sim M^{2j}$.

Let $k>0$  and suppose $1\leq \al< \bt^2$. Then 
$X_{\al\bt}\sim \bt^2\sim M^{2j}$. Moreover, for all $k>0$  
$\tilde\chi_k\neq 0$ implies $\al \geq M^{2j+2k}$.
So $ M^{2j+2k}\leq \alpha < M^{2j}$, which is  impossible.

Let $k=0$ and  suppose $\al> \bt^2\geq 1$. Then 
$X_{\al\bt}\sim \al^{1/4}(\bt^2)^{3/4}$ and  
$(\bt^2)^{3/4}\sim M^{2j}/\al^{1/4}$. Moreover
 $\tilde\chi_0\neq 0$ implies
$1\leq \al \leq M^{2j}$ so  $M^{2j}\geq \al> \bt^2\geq  M^{2j}$. That's 
impossible unless $\al\sim \bt^2$.

Finally let $k=0$ and  suppose $\al> \bt^2$ and $\bt^2\leq 1$. Then 
$X_{\al\bt}\sim \al^{1/4}\sim M^{2j}$.  But $\tilde\chi_0\neq 0$ implies
$1\leq \al \leq M^{2j}$ so $M^{2j}\geq \al\sim M^{8j}$. That's impossible.
The result follows.\qed

\vskip 0.2cm

Now we distinguish three cases. 
\paragraph{Case 1}For $j=0$ we have 
$k_{m}=0$ (no auxiliary scales) and
\begin{equation}\label{c1}
\chi_{jk}(\al,\bt)=\chi_{j} (X_{\alpha \beta })\neq 0 
\quad\Rightarrow \quad \alpha \simeq 1\ , 
\beta^{2}\leq 1 \ .
\end{equation}
\paragraph{Case 2} For $0<j<j_{m}$ we have $k_{m}\geq 1$ and 
$\chi_{jk}(\al,\bt) \neq 0\quad\Rightarrow $ 
\begin{itemize}
 \item{a)\ } $0<k<k_{m}$: then
  \begin{equation}\label{c2a}
    \al \simeq M^{2j+2k}\ , \ |\bt| \simeq M^{j-k/3} 
\end{equation}
\item{b)\ } $ k=0$: then
  \begin{equation}\label{c2b}
  1\leq \al \leq M^{2j}\ , \ |\bt| \simeq M^{j} 
\end{equation}
\item{c)\ } $k=k_{m}$ and $\beta^2\leq 1$: then
 \begin{equation}\label{c2c}
   \alpha\simeq M^{8j}\ , \ 0\leq  \beta^{2}\leq 1 
\end{equation}
\item{d)\ } $k=k_{m}$ and $\beta^2>1$: 
then $k_m=j_{m}-j$, which means $ 4j\geq j_{m}$ and
 \begin{equation}\label{c2d}
   M^{2j_{m}} \leq \alpha \leq M^{8j} \ , \ \beta^{2} \simeq 
\frac{M^{8j/3}}{\alpha^{1/3}}
\end{equation}
Note that in this last case we have $1\leq |\beta |\leq M^{(4j-j_{m})/3}$
\end{itemize}
\paragraph{Case 3}
Finally for $j=j_{m}$ we have $k_{m}=0$ and $\chi_{j_{m}0}(\al,\bt)  \neq 0 
\quad\Rightarrow$ one of the following three situations holds:
\begin{equation}
\begin{array}{ll}
a)\ & M^{8j_{m}} < \alpha<\infty  \ , \ 0\leq \beta^{2}\leq 1\\
b)\ &M^{2j_{m}} \leq \alpha \leq M^{8j_{m}} \ , \ 
\frac{M^{8j_{m}/3}}{\alpha^{1/3} }
\leq  \beta^{2} \leq \alpha   \\
c)\ & 1\leq \alpha \leq \beta^{2} \ , \  M^{2j_{m}}\leq \beta^{2}<\infty   \\
\end{array}
\end{equation}

The typical situation is Case 2a and 2b, that is $0<j<j_{m}$ and 
$0\leq k<k_{m}$.

\section{Scaled Decay}

\begin{theor} \label{propslicdecay}
Let $C^{jk}$ be the scaled propagator introduced in \eqref{subsli}. Then for
any $j< j_m, 0\leq  k\leq k_{m}$ and  $p>0$ the decay is 
\begin{equation}
|C^{jk}(t,x)|\le\ K_p \ 
\frac{M^{-2j-2k/3}}{ [( 1 + f(t) M^{-j-k})(1+|x|M^{-j+k/3})]^p}
\label{eq:scd}\end{equation}
where  $K_{p}$ is a constant dependent on $p$. For the last scale $j=j_m$
we have no decay in $t$ at all and $x$ decay according to the infrared 
cutoff $T$ 
\begin{equation}
|C^{j_m0}(t,x)|\le\ K_p \ 
\frac{T^{2}}{ (1+|x|T)^p}= K_p \ 
\frac{M^{-2j_m}}{ (1+|x|M^{-j_m})^p}
\label{eq:scd1}\end{equation}
\end{theor}

Remark that this decay is identical to the one of \cite{DMR}, section II.5
(equations II.43, II.45 and II.47)

\medskip
\noindent{\bf Proof}
\medskip
The rest of this section is devoted to the proof.
We treat separately the cases listed in sect 3.2.3.

\subsection{Case 1: $\alpha\simeq 1$,  $\beta^2\leq 1$  } 

This corresponds to ${j=k=0}$ so   we need  to prove 
\begin{equation}
|C^{00}(t,x)|\le\ K_p \ 
\frac{1}{ [( 1 + f(t))(1+|x|)]^p}\ .
\end{equation}
Since  $\alpha\simeq 1$ $I$ and $\tilde{I}$ are both bounded by
$K\ (1+ f(t))^{-p}$ (using \eqref{I1b}) thus giving the $t$ decay.
To perform the $\beta$ integral we distinguish two cases:

\paragraph{a)} When $|x|>1$ we need to extract the spatial
decay. We apply
\begin{equation}
e^{-i x^2/(4\beta)}= -i \frac{4\beta^2}{x^2}   
\frac{d}{d\beta} e^{-i x^2/(4\beta)}
\end{equation}
We perfom integration by parts in $\beta$ several times.
Then we obtain the decay $|x|^{-2p}$. Now we can insert
absolute values in the $\beta$ integral that is now bounded by
a constant, since $\beta\leq 1$. Note that the additional 
$\beta$ factor we obtain from integration by parts 
ensures the integral over $\beta$ has no divergence in $\beta=0$.
 
\paragraph{b)} When $|x|\leq 1$ we do not need to extract any spatial
decay. We only need to prove the integral over $\beta$ is bounded.
To avoid the $\beta^{-3/2}$ divergence we go back to \eqref{xrep0}
and after performing several times integration by parts on 
 \begin{equation}
e^{i \beta(k^2-1)}= \frac{1}{1+i(k^2-1)  }   
\left ( 1+\frac{d}{d\beta}\right ) e^{i \beta(k^2-1)}
\end{equation}
we can insert absolute values in the $k$ and $\beta$ integral.
Then no divergence in $\beta$ appears.

\subsection{Case 2a: $\alpha\simeq M^{2j+2k}$, $|\beta|\simeq M^{j-k/3}$ }

This corresponds to  $\,\mb{0\!<\!j\!<\!j_{m}}$,$\,\mb{0\!<\!k\!<\!k_{m}}$.
The $\beta $ integral is performed through 
a saddle analysis. The saddle point for the phase factor $B$ \eqref{Bdef}
is $\bt_s = \pm |x|/2$. Therefore  we introduce a smooth decomposition 
\[
1=\eta(Y) + (1- \eta(Y)   )\ , \qquad  Y=\frac{(\beta-\bt_s)}{\beta_s}\ , 
\qquad
\bt_s = \pm\frac{|x|}{2}\ , \quad x\neq 0
\]
where $\eta$ has compact support $\eta \neq 0$ if 
$Y<1/10$. As 
\[
\partial_{\beta }B (\beta ,x) = 
1 - \frac{1}{\left (1 +\frac{\beta-\bt_s}{\bt_s}\right )^2 }\ , 
\]
it is not difficult to see that inside the support of $1-\eta$
we have  $ |\partial_{\beta }B (\beta ,x) |\ge 1/200$.

\subsubsection{Saddle region} 

This region comes into play only 
when the support of $\eta$ has a non empty interaction with the 
support $\chi^{jk}$ (in this case that means $ |x|\simeq M^{j-k/3}$).
Then we have to study
\begin{equation}
I_s =  
 \int_1^{\infty} d\alpha   \int_{\mathbb{R}}   d\beta \  \chi_{jk}(\al,\bt)
\ \eta\left ( \tfrac{\beta-\bt_s}{\bt_s} \right)\ 
\ 
\frac{{\tilde I} (\alpha,t ) +i\bt I (\alpha ,t) }{\alpha^2 \beta^{3/2}}   \  
e^{- \frac{\beta^2}{4 \alpha}}  e^{-iB (\beta,x)}.
\end{equation}
Near the positive saddle $\bt_s=|x|/2$ we have
\begin{eqnarray}
B(\bt) &=& B(\bt_s) +  \frac{2}{|x|}\ \big(\beta-\bt_s\big)^{2}+  
O\left ( \frac{(\bt-\bt_s)^3}{|x|^2 }\right )\\ \nonumber
& =&B(\bt_s) +  \big(\beta-\bt_s\big)^{2} \frac{2}{|x|}\ 
\left [1+ R(\bt-\bt_s) \right ]\ , \\ \nonumber 
R(\bt-\bt_s)  &\propto & \frac{|\bt-\bt_s|}{|x|}< 1/10 .
\end{eqnarray}
As we have a phase factor, it is not easy to perform the Gaussian integral
in $y = \bt-\bt_s$, but we know the result should be 
$\sqrt{|x|}= M^{(j-k/3)/2}$.
In order to prove that $|y|\leq \sqrt{|x|}$ we perform integration by parts in
the following way: 
\begin{equation}
e^{-iB (\beta ,x)} =   \frac{1}{(1-i\ \sqrt{x}\ \partial_{\beta }B (\beta,x))}
\left ( 1 + \sqrt{x} \frac{\partial}{\partial\bt}\right ) e^{-iB (\beta ,x)}\ . 
\end{equation}
As $-iB$ is a phase factor, $-i \partial_{\beta }B$ is pure imaginary 
so the denominator is well defined. 
Now for $|y| = |\bt-\bt_s| \leq |x|$ we have
\[
\partial_{\beta }B (\beta ,x) \ =\ 4  \frac{y}{|x|} \ [ 1 + R(y)] \qquad 
\mb{ \rm with}\quad 
 |R(y)| 
\propto \frac{|y|}{|x|} < 1/10\ .
\]
Therefore $|\partial_{\beta }B (\beta ,x) | \geq   \frac{|y|}{|x|}$ and
\begin{equation}
   \left |\frac{1}{(1+\ \sqrt{x}\partial_{\beta } B)}\right |\leq  
    \frac{1}{(1+\frac{|y|}{\sqrt{x}})}. 
\label{eq:yb}\end{equation}
Performing integration by parts with respect to $\beta $ we get 
\begin{align}
I_s = & 
 \int_1^{\infty} d\alpha   \int_{\mathbb{R}}   d\beta \ 
\frac{ e^{- \frac{\beta^2}{4 \alpha}}  e^{-iB (\beta ,x)}  }{ 
(1-i\ \sqrt{x}\ \partial_{\beta }B)}
 \frac{{\tilde I} (\alpha ,t) +i\bt I (\alpha ,t) }{\alpha^2 \beta^{3/2}} \\
 & \hskip 4cm
\left [ \chi_{jk}(\al,\bt)
\ \eta\left ( \tfrac{\beta-\bt_s}{\bt_s} \right) 
\left( 1- i\tfrac{|x|\partial_{\beta }^{2}B}{ (1-i\sqrt{|x|}\partial_{\beta }B 
)}\right) + R(\alpha,\beta)
  \right ]\nonumber
\end{align}
where 
\begin{equation}
|R(\alpha,\beta)|\leq  O\left (  \tfrac{\sqrt{|x|}}{\bt}\right )  
 \bar\chi_{jk}(\al,\bt)
\ \bar\eta\left ( \tfrac{\beta-\bt_s}{\bt_s} \right) 
\end{equation}
and $\bar\chi_{jk}$, 
$\bar\eta$  have a slightly larger support than 
$\chi_{jk}$, $\eta$.

Repeating once and inserting absolute values inside the integral
we obtain
\begin{equation}
|I_s| \leq 
 \int_1^{\infty} \frac{d\alpha}{\alpha^2}   \int_{\mathbb{R}}   
\frac{d\beta}{|\beta|^{3/2}} \ 
 \tfrac{ \bar\chi_{jk}(\al,\bt)
\  \bar\eta\left ( \tfrac{\beta-\beta_s}{\beta_s} \right)}{
\left (1+\frac{|\bt-\bt_s|}{\sqrt{|x|}}\right  )^2}  
 e^{- \frac{\beta^2}{4 \alpha}} 
\ (|{\tilde I}|+ |\bt| |I|) 
\label{eq:bsaddle}\end{equation}
where we applied \eqref{eq:yb} and $\sqrt{|x|}/\beta<1$.
We bound  $\beta\ e^{-\frac{\beta^2}{4\alpha}} 
\leq \sqrt{\alpha} e^{-\frac{\beta^2}{4\alpha}}$.  Using \eqref{I1b}
to bound $I$ and $|\tilde I|$ and $\alpha \simeq M^{j+k}$ we  get
\begin{eqnarray}
|I_s| &\leq & \frac{K_p}{(1+ f(t) M^{-j-k})^p}
 \int_1^{\frac{1}{T}} \frac{d\alpha}{\alpha^2} \sqrt{\alpha} 
\int_{-\infty}^\infty   
\frac{d\beta}{\beta_s^{3/2}} \ 
 \frac{ \bar\chi_{jk}(\al,\bt_s)
\  \bar\eta\left ( \tfrac{\beta-\beta_s}{\beta_s} \right)}{
\left (1+\frac{|\bt-\bt_s|}{\sqrt{|x|}}\right  )^2}   
e^{-\frac{\bt^2_s}{4\al}}   
\cr
& \leq & \frac{K_p}{(1+ f(t) M^{-j-k})^p}\int_{\al \simeq M^{2j+2k}}
\frac{d\al}{\al^{3/2}}   
\int_{-\infty}^{\infty} \frac{dy}{\left (1+\frac{|y|}{\sqrt{|x|}}\right  )^2 }  
\frac{1}{|x|^{3/2}} 
\cr 
& \leq & \frac{K_p}{(1+ f(t) M^{-j-k})^p}   \frac{M^{2j+2k}}{M^{3j+3k}}
\frac{\sqrt{|x|}}{|x|^{3/2}}
\cr
& =&\frac{K_p}{(1+ f(t) M^{-j-k})^p}
M^{-2j - \frac{2}{3} k} .
\end{eqnarray}
In the first line $\bar\eta$ ensures $\beta\simeq \beta_s=|x|\simeq M^{j-k/3}$.
As $|x|\simeq M^{j-k/3}$ we do not need to gain any further  spatial decay
hence Lemma \ref{propslicdecay} holds in this case.

\subsubsection{Region far from the saddle}

This region comes into play only 
when the support of $1-\eta$ has a non empty interaction with the 
support $\chi^{jk}$. 
Then we have to study
\begin{equation}
I_f =  
 \int_1^{\infty} d\alpha   \int_{\mathbb{R}}   d\beta \  \chi_{jk}(\al,\bt)
\ \left [ 1- \eta\left ( \tfrac{\beta-\bt_s}{\bt_s} \right)\right ]\ 
\ 
\frac{({\tilde I} (\alpha ,t) +i\bt I (\alpha ,t)) }{\alpha^2 \beta^{3/2}}  \ 
e^{- \frac{\beta^2}{4 \alpha}}  e^{-iB (\beta ,x)} .
\end{equation}
In this region $\partial_{\beta }B>1/200$ so we can apply
\begin{equation}
e^{-iB (\beta ,x)} = \frac{1}{\partial_{\beta }B}  
\frac{\partial}{\partial\bt} e^{-iB}\ .
\end{equation}
Performing integration by parts $p+1$ times and inserting absolute values
 we get
\begin{equation}\label{outsaddle1}
|I_f|\leq K_p \int d\al \int d\bt  \  \bar\chi_{jk} (1-\bar \eta_1) \  
e^{- \frac{\beta^2}{4 \alpha}} e^{-iB} 
\frac{(|{\tilde I}|+ |\bt| |I|) }{\alpha^2 |\beta|^{3/2}} 
\frac{1}{ |\partial_{\beta }B|^{p+1}}\  O\left ( \frac{1}{|\bt|^{p+1}}\right ) .
\end{equation}
where $\bar \eta_1$ has slightly smaller support than $\eta$ (and $\bar\chi$
is the same as in the saddle region)
Note that when the derivative hits $\eta$ instead of a $1/\beta$
we get a $1/\beta_s$ factor. But 
\begin{equation}
\frac{1}{\bt_s} \eta'  \le const (1- \bar \eta_1 )
\frac{1}{\vert \bt \vert } .
\end{equation}Since $\partial_{\beta  }B = 1 - |x|^2/ 4\bt^2$, 
the $1-\bar \eta_1$ function ensures that 
$\vert \vert 2 \bt /x \vert - 1\vert \ge const $, and 
$\vert \bt \vert  \simeq M^{j-k/3}\ge 1$ we have
\begin{equation}
\frac{1}{ |\partial_{\beta }B|^{p+1}}\leq \frac{K}{(1+|x|M^{-j-k/3})^p }
\end{equation}
The factor $\beta^{-1-p}$ ensures the global factor is correct. Actually
we need only to use $\beta^{-1}$:
\begin{align}\label{outsaddle2}
|I_f| &\leq \ \frac{K_p}{(1+ |x|^2/M^{j-k/3})^p} \frac{1}{(1+ f(t)M^{-j-k})^p} 
\ \int 
\frac{d\al}{\al^{3/2}}   \int \frac{d\bt}{\bt^{5/2}}   \bar\chi_{jk} \cr
&\ \leq \   \frac{K_p}{(1+ |x|^2/M^{j-k/3})^p}  \frac{1}{(1+ f(t)M^{-j-k})^p}
\ M^{-j - k}  M^{-(3/2)(j-k/3)}\cr
 & \leq \   \frac{K_p}{(1+ |x|^2/M^{j-k/3})^p}  \frac{1}{(1+ f(t)M^{-j-k})^p}
\   M^{-2j - \frac{2}{3} k} 
\end{align}
as $j-k/3 \geq 0$.

\subsection{Case 2b: $1\leq \alpha \leq M^{2j}, |\beta|\simeq M^{j}$}
This corresponds to $k=0, 0<\! j\!<\!j_{m}$ so  we need to prove
\begin{equation}
|C^{j0}(t,x)|\le\ K_p \ 
\frac{M^{-2j}}{ [( 1 + f(t) M^{-j})(1+|x|M^{-j})]^p}
\end{equation}
This case is treated exactly as the $k>0$ case. 
The only difference is that now
$\al$ has no fixed value, but instead must be integrated between 
1 and $\bt^2$. 
This can be done using the exponential decay (after performing
all the necessary integration by parts in the $\beta$ integral)
\begin{equation} \label{equatcase3}
\int_1^{\bt^2} \frac{d\al}{\al} \frac{1}{\sqrt{\al}} e^{-\frac{\bt^2}{\al}}
= \frac{1}{\bt}  \int_{\bt^{-2}}^{1} \frac{d\al}{\al} \frac{1}{\sqrt{\al}} 
e^{-\frac{1}{\al}}
\leq \ \frac{K}{\bt} \simeq \ \frac{K}{M^j}
\end{equation}
where $K$ is some constant.
As a consequence the $\al$ integral is bounded by $M^{-j}$. 
The $t$ decay can be obtained observing that as $\al\leq \bt^2$ then
\begin{equation}
\frac{1}{\left(1+ \frac{f(t)}{\sqrt{\al}} \right)^p}  \leq 
\frac{1}{\left(1+ \frac{f(t)}{\bt} \right)^p} \simeq  
\frac{1}{\left(1+ \frac{f(t)}{M^j} \right)^p} .
\end{equation}
The $x$ decay is treated by the same saddle/offsaddle analysis as 
for $k > 0$.
 
\subsection{Case 2c: $\alpha\simeq M^{8j}$, $\beta^2\leq  1$}
This corresponds to  $ 0<\! j\!<\!j_{m}$ and  $k=k_m$ so 
we must distinguish between the two possible 
values for $k_m$.

\paragraph{(i)} If $k_m=3j$, then $4j\leq j_m$ and we must prove
\begin{equation}
|C^{jk_m}(t,x)|\le\ K_p \ 
\frac{M^{-4j}}{ [( 1 + f(t) M^{-4j})(1+|x|)]^p}
\end{equation}
Since $\beta^{2}\leq 1 $, we treat the $\beta$ integral (and $|x|$ decay)
as in Case 1 (distinguish $|x|>1$ and $|x|\leq 1$). The bounds
on $I$ and $\tilde I$ from \eqref{I1b}  give the correct decay in $t$ and 
the $\alpha$ integral gives the prefactor $M^{-4j}$. 

\paragraph{(ii)} If $k_m=j_m-j$, then $4j\geq j_m$ and we want to prove
\begin{equation}
|C^{jk}(t,x)|\le\ K_p \ 
\frac{M^{-4j} M^{2/3(4j-j_m)}}{ [(1+f(t)M^{-j_m})(1+|x|M^{-(4j-j_m)/3})^{p}]}
\end{equation}
Since $4j\geq j_m$ we have $\alpha\simeq M^{8j}\geq M^{2j_m}=1/T^2$.
Therefore the bound on $I$, $\tilde I$ from \eqref{I1b} give the correct decay 
in $t$. For $|x|$, repeating the same arguments as in Case 1, it is
easy to get a decay $(1+|x|)^{-p}\leq (1+|x|M^{-(4j-j_m)/3})^{-p}$ 
since $4j-j_m\geq 0$. Finally the $\alpha $ integral is bounded
by $M^{-4j}\leq M^{-4j} M^{2/3(4j-j_m)}$.

\subsection{Case 2d:  $M^{2j_m}\leq \alpha \leq M^{8j}$, 
$\beta^2\simeq  \frac{M^{8j/3}}{\alpha^{1/3}}$  }
This corresponds to $ 0<\! j\!<\! j_{m}$ and $ k=k_m$ with $k_m=j_m-j $. 
and we need to prove
\begin{equation}
|C^{jk_{m}}(t,x)|\le\ K_p \ 
\frac{M^{-4j} M^{2/3(4j-j_m)}}{[(1+|x|M^{-(4j-j_m)/3})^{p}]}
\ .
\end{equation}
Note that in this case there is no $t$ decay.
Since  $M^{2j_m}\leq \alpha$ the bounds \eqref{I2b} on $I$, $\tilde I$ 
give the correct decay in $t$. The $\beta$ integral is performed by
a saddle analysis (region near/far from the saddle) as in Case 2a.

The result is the correct decay in $x$ times
a factor $(\sqrt{\alpha}+\beta_s)\sqrt{\beta_s}/(\alpha^2\beta_s^{3/2})$,
where we used $\beta^2\leq \alpha$. Inserting the value of 
$\beta_s^2= M^{8j/3}/\alpha^{1/3} $ and performing the $\alpha$ integral
we get the correct prefactor.

\subsection{Case 3}
The last case corresponds to $j=j_{m}$, so $k_{m}=0$, 
so we want to prove
\begin{equation}
|C^{j_{m}0}(t,x)|\le\ K_p \ 
\frac{M^{-2j_m} }{ (1+|x|M^{-j_m})^{p}}
\end{equation}
We have three possible ranges for $\alpha $ and $\beta $.

\subsubsection{Case  3.a}  $M^{8j}\leq \alpha \leq  \infty$, $\beta^{2}\leq 1$.
Since $\alpha > M^{2j_{m}}$ we apply \eqref{I2b} so
\[
|I+i\beta \tilde{I}| \leq\ K (T^{-1}+\beta) (1+f (t)T)^{-1}\ .
\]
Since $\beta^{2}\leq 1$, by applying the same $|x|>1$/$|x|\leq 1$ analysis
as in Case 1 we obtain
\[
\int d\beta  \chi_{jk} (\alpha,\beta) F (\alpha,\beta,t, \vec{x} )
\ \leq  \  \frac{1}{\alpha^{2}}
\frac{K_{p} M^{j_{m}}}{ (1+f (t)T)\ ( 1+|x|)^{p} }  \leq  \  
\frac{K_p}{\alpha^{2}}\frac{1}{ ( 1+|x|)^{p} }
\]
The $\alpha $ integral is then bounded by $M^{-8j}$ so
\[
|C^{j_{m}0}(t,x)|\le\ 
\frac{ K_p \ M^{-8j_m} }{  (1+|x|)^{p} }
< \
\frac{ K_p \ M^{-2j_m} }{ (1+|x|M^{-j_m})^p}\ .
\]

\subsubsection{Cases 3.b and 3.c}
To get the $|x|$ decay we use the infrared cutoff on $k_0$ ($|k_0|\geq T$)
(as we did in \cite{DMR} equation II.9). Let consider first the case 
$|x|>1$.

Inside the $\beta$ integral we apply the identity:
\begin{equation}
e^{- \frac{|x|^2}{4\beta}} = \frac{4\beta^2}{|x|^2} 
\frac{\partial}{\partial\beta}
e^{- \frac{|x|^2}{4\beta}}
\end{equation}
Performing integration by parts with respect to $\beta$ $p$ times we obtain
a factor of order 
\begin{equation}
\left (\frac{\beta^2}{|x|^2}\right )^p = \frac{1}{(T|x|)^{2p}} 
\left (\frac{\beta^2}{\alpha}\right )^p (\alpha T^2)^p
\end{equation}
The first term is exactly the decay we are looking for.
The integral over $\beta$ is now performed using the saddle analysis.
The factor $(\beta^2/\alpha)^p$ is bounded using a piece
of the exponential decay $e^{- \frac{\beta^2}{4 \alpha}}$ (after 
performing the necessary integrations by parts in the $\beta$ integral).
The factor $ (\alpha T^2)^p$ is bounded by the $e^{-k_0^2\alpha}$
in $I$, $\tilde I$:
\begin{eqnarray}
 (\alpha T^2)^p |I| &\leq & \sqrt{\al} T 
\sum_{k_0}  \ (\alpha T^2)^p \  e^{-k_0^2 \alpha} \ \leq\   c_p \\
 (\alpha T^2)^p |\tilde I| &\leq & \sqrt{\al} T 
\sum_{k_0} \ (\alpha T^2)^p \ 2\alpha |k_0|\   e^{-k_0^2 \alpha} \  \leq\   
\sqrt{\alpha} c'_p \\
\end{eqnarray}
where we used the infrared cutoff $|k_0|\geq T$.

The saddle analysis is performed as in Case 2a 
(see \eqref{eq:bsaddle}-\eqref{outsaddle1})
\begin{align}
|I_s| & \leq  \frac{K_p}{(T|x|)^{2p}} 
 \int \frac{d\alpha}{\alpha^2}   \int  
\frac{d\beta}{|\beta|^{3/2}} \ 
 \tfrac{ \bar\chi_{j_m0}(\al,\bt)
\  \bar\eta\left ( \tfrac{\beta-\beta_s}{\beta_s} \right)}{
\left (1+\frac{|\bt-\bt_s|}{\sqrt{|x|}}\right  )^2}  
\ (\sqrt{\alpha}+ |\bt|)
\ e^{- \frac{\beta^2}{8 \alpha}} \cr
|I_f|& \leq  \frac{K_p }{(T|x|)^{2p}}  
\int\frac{d\alpha}{\alpha^2}  \int \frac{d\beta}{|\beta|^{3/2}}  \
 \tfrac{ \bar\chi_{jk} (1-\bar \eta_1)}{|\beta|} \  
(\sqrt{\alpha}+ |\bt|) \  e^{- \frac{\beta^2}{8 \alpha}} .
\end{align}
where we bounded $|B'|^{-1}\leq const$ since we already have the $x$
decay we need.
Since $I_f$ is done in the same way as $I_s$ (and the bounds are easier) 
we will only look at $I_s$. This last appears only when $|x|$ belongs
to the integration interval for $\beta$. 

\paragraph{(b)} In this case we first perform the $\beta$ integral.
Since $\beta\leq \sqrt{\alpha}$ we have
$(\sqrt{\alpha}+ |\bt|)\leq 2 \sqrt{\alpha}$. 
The result is $\sqrt{|x|}/|x|^{3/2}$, then 
\begin{equation}
|I_s|  \leq  \frac{K_p}{(T|x|)^{2p}} 
 \int \frac{d\alpha}{\alpha^2} \frac{\sqrt{\alpha}}{|x|}
\leq 
\frac{K_p}{(T|x|)^{2p}} 
 \int_{M^{2j_m}}^{M^{8j_m}} 
\frac{d\alpha}{\alpha^{3/2}} \frac{\alpha^{1/6}}{M^{4j_m/3}}
\leq K_p \frac{M^{-2j_m}}{(T|x|)^{2p}} 
\end{equation}
\paragraph{(c)} In this case we first perform the $\alpha$ integral.
Since $\beta\geq \sqrt{\alpha}$ we have
$(\sqrt{\alpha}+ |\bt|)\leq 2 |\beta|$. 
We perform the $\alpha$  integral as in \eqref{equatcase3}:
\begin{equation}
\int_1^{\beta^2} \frac{d\alpha}{\alpha^2} e^{-\beta^2/\alpha} \leq 
\frac{K}{\beta^2}\ .
\end{equation}
The $\beta$ integral is then bounded by
\begin{equation}
|I_s|  \leq  \frac{K_p}{(T|x|)^{2p}}
\frac{|x|}{|x|^2} \frac{|x|^{1/2}}{|x|^{3/2}}
\leq  \frac{K_p}{(T|x|)^{2p}} \frac{1}{|x|^2} 
\leq K_p \frac{M^{-2j_m}}{(T|x|)^{2p}} 
\end{equation}
since $|x|\geq M^{j_m}$ (we are in the saddle region).

Remark that when $|x|\leq 1$ we can repeat the bounds above 
without extracting any $x$ decay.  \qed

\end{document}